\documentclass[twocolumn,showpacs,preprintnumbers,amsmath,amssymb,aps]{revtex4}
\input epsf
\usepackage{graphicx}
\begin{document}

\title{Fluctuations, Jamming, and Yielding 
for a Driven Probe Particle in Disordered Disk Assemblies} 
\author{C.J. Olson Reichhardt and C. Reichhardt} 
\affiliation{ 
Theoretical Division, Los Alamos National Laboratory, Los Alamos, New Mexico 87545
}

\date{\today}
\begin{abstract}
Using numerical simulations we
examine the velocity fluctuations of a probe particle driven with
constant force through 
a two-dimensional disordered assembly of 
disks which has a well defined jamming point J at
a density of $\phi_{J} = 0.843$. 
As $\phi$ increases toward $\phi_J$,
the average velocity of the probe particle decreases  
and the velocity fluctuations show 
an increasingly intermittent or avalanchelike behavior.
When the system is within a few percent of the jamming density, the
velocity distributions are exponential, while when the system is
less than a percent away from jamming, the velocity distributions have a
non-exponential or power law character.
The velocity power spectra
exhibit a crossover from a Lorentzian form to a $1/f$ shape near jamming.     
We extract a correlation length exponent $\nu$ which is in good
agreement with recent shear simulations.     
For $\phi > \phi_{J}$, there is a critical threshold force $F_{c}$
that must be applied for the probe particle to 
move through the sample
which increases with increasing $\phi$.  
The onset of the probe motion above $\phi_{J}$ occurs 
via a local yielding of 
the particles around the probe particle which we term  
a local shear banding effect. 
\end{abstract}
\maketitle

\vskip2pc

\section{Introduction}
When a disordered assembly of particles with short-range steric interactions
becomes dense enough that the particles begin to overlap, a transition occurs
from an unjammed state where the system easily flows or acts 
like a liquid to a jammed state where the system resists shear and shows
a solid-like response.
Liu and Nagel proposed
a general phase diagram for disordered particle assemblies 
where a transition into a jammed state occurs as a function of
several different 
variables including increasing density, decreasing external forcing, or
decreasing temperature \cite{Nagel}. 
At `Point J' on this phase diagram,
jamming occurs for increasing density at zero stress and zero temperature
\cite{Hern}.
There is growing evidence that point J has the properties
of a critical point with diverging length scales on both sides
of the jamming transition 
\cite{Hern,Drocco,Silbert,Silke,Wyart1,S,El,B,Olsson,Wyart,Barrat,Hatano,Head,Zhang}.   
Point J has been well studied in the particular system of
two-dimensional, finite range repulsive frictionless bidisperse disks
with a radii ratio of $1:1.4$ 
\cite{Hern,Drocco,Olsson,Barrat,Head}. 
The bidispersity 
is necessary to prevent the system from crystallizing. 
Simulations show that the jamming density $\phi_{J} \approx 0.84$, and   
experimental realizations of this
system also give a similar jamming density \cite{Zhang,CD}. 
    
One aspect of jamming that has been less well studied is the 
dynamical response as point J is approached. 
If point J is a critical point, 
it could have strong effects on the dynamical response of the system to global
or local forcing and may give rise to 
intermittent or avalanche behavior. 
In the numerical work of Drocco {\it et al.} \cite{Drocco}, 
the velocity fluctuations of a single probe particle driven with constant force
showed signs of increasing intermittency and avalanche characteristics as
$\phi$ approached $\phi_J$.
More recent experimental work in which
a single probe particle was driven at constant force through
a granular packing
also indicates that the velocity fluctuations become 
increasingly intermittent as
$\phi_{J}$ is approached \cite{CD}.   
These results suggest that proximity to point J will strongly affect
the dynamics and may result in the appearance of critical properties.

The velocity and force characteristics of a
locally driven particle have already been used 
as a probe of the yielding and dynamical properties of 
a two-dimensional disordered colloid system, 
where
a finite threshold force must be applied 
before the particle can be set in motion 
\cite{Hastings,LM}.
Experiments and simulations of three-dimensional disordered colloidal systems 
have shown that a
threshold force for motion appears when the density is high enough for a  
glass transition to occur \cite{Weeks,Ga}, while
for jamming in two dimensions a finite threshold force is also expected
to appear above the jamming transition.
Drag effects in granular systems have been studied by moving a large
obstacle through the granular media
\cite{Albert,Albert2}. 
Geng and Behringer analyzed the fluctuations of a 
probe particle held fixed in a two-dimensional
system while the remaining grains were driven past \cite{Geng}, and found that
the force fluctuations have intermittent characteristics and are
exponentially distributed.

In this work we examine the velocity fluctuations of a probe disk driven
with constant force 
through a two-dimensional bidisperse system of disks as the density of 
the system passes through point J.
As $\phi_{J}$ is approached from below,
the velocity fluctuations show increasingly intermittent properties 
characterized by large bursts of motion.      
The average probe particle velocity 
initially decreases linearly with increasing $\phi$; however, within 
a few percent of the jamming density, the average velocity falls to zero
in a nonlinear manner.  
The velocity fluctuations are exponentially distributed
when $\phi$ is within several percent of the jamming density, while
within one percent of $\phi_{J}$ the velocity fluctuations 
have a non-exponential distribution which can be  
fit with a power law. 
The power spectra shows a Lorentzian behavior with 
a $1/\omega^{2}$ behavior at high frequencies that is similar to the results found
in the slow drag experiments of Ref.~\cite{Geng}; however, very close to jamming
the low frequency response exhibits a $1/\omega$ behavior. 
The local rearrangements of disks near the probe particle
increase in spatial extent as the jamming transition is approached from below. 

For $\phi < \phi_{J}$, the probe particle is
always able to move through the sample for 
arbitrarily small driving forces.  In contrast, for
$\phi > \phi_{J}$, there is a finite threshold force $F_c$ required to
move the probe particle.
Above the threshold force, progress of the probe particle is made possible
by yielding of the surrounding particles that occurs via local plastic 
rearrangements of the disks near the moving probe particle.
The appearance of a finite driving threshold in the jammed phase 
is similar to recent numerical studies of dense colloidal  
systems where a threshold force appeared in sufficiently dense
samples \cite{Ga}.  We find that the
threshold force increases with increasing $\phi$.     
The velocity fluctuations at drives just above the yielding transition
$F_c$ have a nonexponential distribution,
while at higher drives the velocity fluctuations 
become exponentially distributed.
When the probe particle moves, it creates local rearrangements or a 
local shear banding.     

\section{Simulation}
We simulate a two-dimensional bidisperse system of $N$ disks of 
size $L \times L$ with
periodic boundary conditions in the $x$ and $y$ directions.
The particles are modeled as repulsively harmonically interacting disks 
of radii $R_{A}$ or $R_{B}$ with a size ratio $R_A:R_B$ of $1.4:1$. 
The ratio of A and B particles is $50:50$.  
The initial disk packing is obtained
by placing the disks in non-overlapping locations and then 
shrinking all disks, adding a few additional disks, and reexpanding all disks
while thermally agitating the disks. 
This process is repeated until the desired density is obtained.
After the disk configuration is generated, we select a single 
disk and drive it with a constant force ${\bf F}_{D}=F_D{\bf {\hat x}}$. 
The motion of an individual disk $i$ at position ${\bf r}_i$ is 
governed by the following overdamped equation:
$$\eta \frac{d{\bf r}_{i}}{dt} = \sum_{i\neq j}k(R_{eff}-|{\bf r}_{ij}|)
\frac{{\bf r}_{ij}}{|{\bf r}_{ij}|}
\Theta (R_{eff}-|{\bf r}_{ij}|) + {\bf F}_{D}^i
$$
Here $\Theta$ is the Heaviside step function,
${\bf r}_{ij} = {\bf r}_{i} - {\bf  r}_{j}$, and
$R_{eff}= R_i+R_j$, where $R_i$ and $R_j$ are either $R_A$ or $R_B$ depending
on the type of each disk.
We take the spring constant $k = 200$.
The driving term ${\bf F}^i_{D}=F_D{\bf {\hat x}}$ is applied to only one disk and we can 
extract the instantaneous velocities of all disks.
In this work, $L = 60$ and the jamming density is $\phi_J\approx 0.843$, 
corresponding to approximately $N=2600$ disks. 
The ability to 
experimentally realize a system  where the probe particle velocity can be 
examined under a constant driving 
force has recently been demonstrated \cite{CD}.  

\section{Velocity Fluctuations Below Jamming} 

\begin{figure}
\includegraphics[width=\columnwidth]{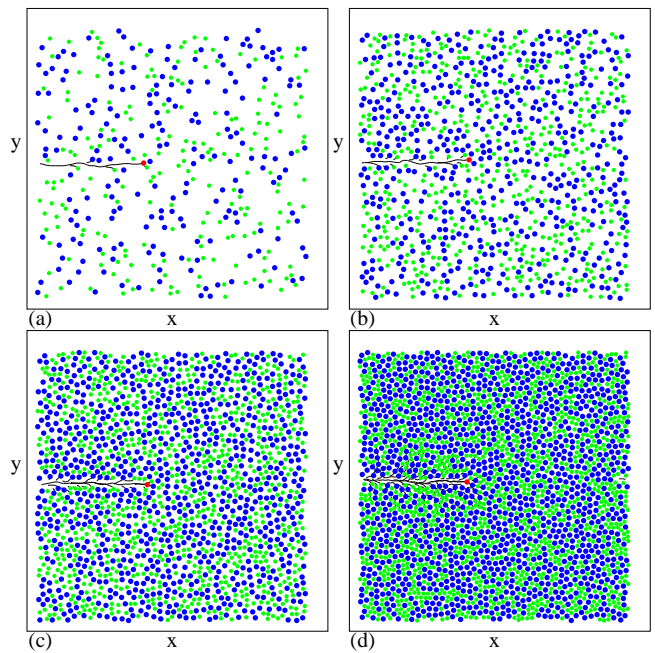}
\caption{
Images of the system at different densities where the probe particle (red) 
is driven $2/5$ of the way through a disordered sample of bidisperse
disks.
The smaller disks are light colored (green) and the larger disks are 
dark colored (blue). 
Black lines indicate the trajectories of the disks that have moved.
(a) $\phi = 0.13$, (b) $\phi = 0.323$,
(c) $\phi = 0.55$, and 
(d) $\phi = 0.81$. 
}
\label{fig_schematics}
\end{figure}

In Fig.~1 we show snapshots of the system at different densities
of $\phi = 0.13$, 0.323, 0.55, and $0.81$. 
The probe particle is highlighted and the trajectories of all grains are
indicated as the probe moves $2/5$ of the way through the sample.
The larger disks are dark colored, the smaller disks are light colored, 
and the overall disk arrangement is disordered. 
At each of the densities presented, the probe particle leaves an empty space
behind it after pushing the other disks out of the way, as is most evident
in Fig.~1(c).
The appearance of an empty trail behind the probe
particle generally occurs for any density below $\phi_{J}$ 
since there is no pressure in the 
system to force the surrounding particles back into the empty space. 

\begin{figure}
\includegraphics[width=\columnwidth]{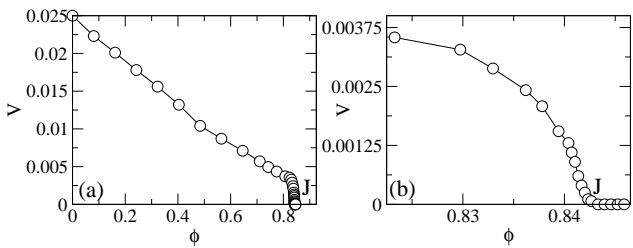}
\caption{
(a) The average velocity $\langle V\rangle$ 
of the probe particle vs $\phi$ for $F_{D} = 0.025$. 
At low densities, $\langle V\rangle$ varies linearly with $\phi$, and
a small change in the slope appears near $\phi = 0.5$.
$\langle V\rangle$ goes to zero at the jamming density $\phi_{J} = 0.843$,
indicated by the letter J. 
(b) A detail of $\langle V\rangle$ vs $\phi$
from (a) showing that $\langle V\rangle$ does not start to fall  
rapidly until $\phi$ is within
three percent of the jamming density.     
}
\label{fig_grainboundaries}
\end{figure}

To construct velocity histograms for different densities at fixed
driving force, we    
perform a series of simulations at increasing values of $\phi$ and measure the
instantaneous time trace of the probe particle velocity in the direction of
the drive,
$V(t)$. 
We then average the velocities over the entire trace to obtain 
$\langle V\rangle$.  
In Fig.~2(a) we plot $\langle V\rangle$
vs $\phi$ for a system with $F_{D} = 0.025$. Here $\langle V\rangle$  
monotonically decreases with  $\phi$ and goes to zero at 
$\phi_{J} = 0.843$, a jamming density which is
in agreement with 
other studies \cite{Barrat,Head}.
Fig.~2(a) also shows that slope of the decrease
in $\langle V\rangle$ with increasing $\phi$ 
becomes shallower above $\phi = 0.5$, and that $\langle V\rangle$ changes
from a roughly linear decrease with increasing $\phi$ to a much more
rapid nonlinear decrease
for $\phi > 0.82$. 
When $\phi < 0.5$, the probe particle motion alternates
between free motion unimpeded by any disks and occasional collisions during
which the probe particle pushes one additional disk along with it.
We show later that this appears as  
a clear feature in the velocity distributions. 
For $\phi > 0.5$, the probe particle spends most of its time pushing
one or more particles but the number of dragged particles
remains small.  
The distance over which the probe particle drags 
individual background disks varies widely due to interactions among the 
dragged disks which create random fluctuations for the dragged disk 
closest to the probe particle.
For $\phi > 0.82$, the induced motions in the background packing 
become increasingly collective, and even disks which are far away from the
probe particle can participate in the motion.
In Fig.~2(b) we plot $\langle V\rangle$ versus $\phi$ 
close to jamming.
The velocity decreases in a nonlinear fashion within three percent
of the jamming density.  Within less than one percent of $\phi_{J}$, the 
velocity curve is still nonlinear; however, the concavity of the slope
of $\langle V\rangle$ versus $\phi$ changes sign. 
In a later section we discuss how the  
velocity scales near jamming and 
how it can be related to a diverging correlation length.  

\begin{figure}
\includegraphics[width=\columnwidth]{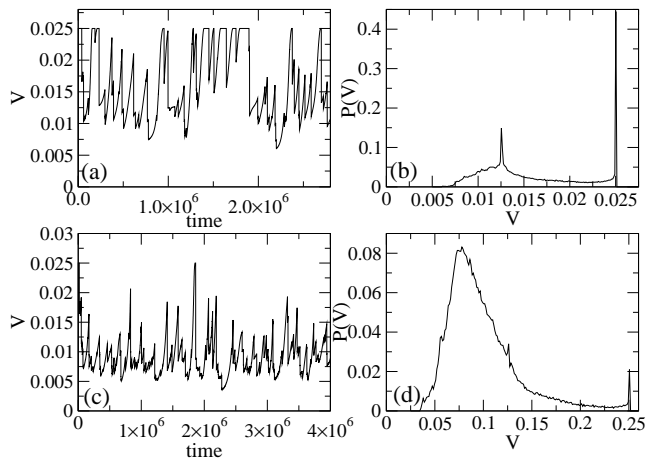}
\caption{
(a) $V(t)$, the time trace of the probe particle velocity, at $\phi = 0.323$
for $F_{D} = 0.025$. Between collisions the probe particle moves at 
a velocity of $V=0.025$ due to the overdamped dynamics.
(b) The histogram $P(V)$ of the velocities for $\phi = 0.323$ for 
ten different initial conditions at $F_{D} = 0.025$. The peak at 
$V = 0.025$ corresponds to the collision-free portion of the
particle motion, while the second peak near $V=0.0125$ corresponds to the
times that the probe particle is pushing one extra particle which
doubles the drag.  All of the $P(V)$ in this work are normalized such that 
the probabilities sum to one.  
(c) $V(t)$ at $\phi = 0.55$ and
$F_{D} = 0.025$. (d) $P(V)$ for several realizations at $\phi=0.55$ and
$F_D=0.025$.
}
\label{fig_plots}
\end{figure}

We next examine in detail the velocity distributions obtained from
time series $V(t)$ taken at 
fixed values of $\phi$.
In Fig.~3(a) we plot $V(t)$ for a sample with $\phi = 0.323$
and $F_D=0.025$, 
the same system shown in Fig.~1(b).
This single time trace is taken during the period required for the particle
to move slightly more than halfway through the sample.
Figure~3(a) shows that the probe particle often moves 
at the maximum possible velocity $V=0.025$.  The plateaus at $V=0.025$
occur when the probe particle is not in contact with any other disks. 
Collisions with other disks reduce the velocity of the probe particle
and in some cases the probe pushes one or more background particles 
for a period of time, creating 
characteristic drops in $V(t)$ of $50\%$ or
more as seen in Fig.~3(a).  
In Fig.~3(b) we plot the distribution of the
velocities $P(V)$ for the system in Fig.~3(a).  
The histogram is obtained 
from ten simulations in which
the probe particle is selected at different locations in the sample
and driven over equal distances.
In this work all $P(V)$ curves are normalized such that
the sum of the probabilities 
equals $1.0$. In Fig.~3(b) 
a large peak appears at $V = 0.025$, corresponding to time periods
when the probe particle moves freely without colliding with any background
particles, as indicated by the
upper plateaus at $V = 0.025$ in $V(t)$ in Fig.~3(a). 
Figure 3(b) shows that there is a  second 
peak in $P(V)$ centered at $V=0.0125$, 
which is half of the value of the free particle velocity. 
This peak corresponds to the time 
the probe particle spends pushing 
one additional disk, which doubles the
drag. The same double peak feature in $P(V)$ appears for all
$\phi < 0.5$.  In this regime, we find that as $\phi$ increases,
the relative weight of the $V=0.025$ peak decreases and is shifted to
the $V=0.0125$ peak.
In Fig.~3(c,d) we plot $V(t)$ and $P(V)$
for a higher $\phi = 0.55$.  Here the weight at $V = 0.025$ is
significantly reduced and the peak probability has shifted to 
$V = 0.008$. 
There is still a peak at $ V = 0.0125$ 
which is now higher than than the $V=0.025$ peak.  

\begin{figure}
\includegraphics[width=\columnwidth]{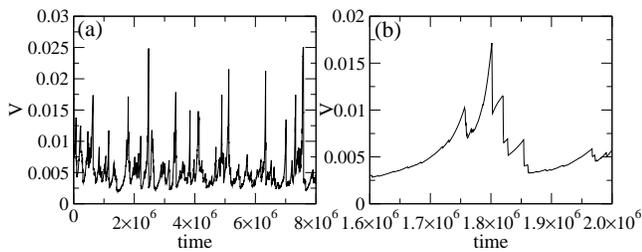}
\caption{
$V(t)$
for $\phi = 0.71$ and $F_D=0.025$, showing a more 
intermittent motion. (b) A blowup of one of the velocity pulses indicating
an asymmetry in the response. 
 }
\end{figure}

\begin{figure}
\includegraphics[width=\columnwidth]{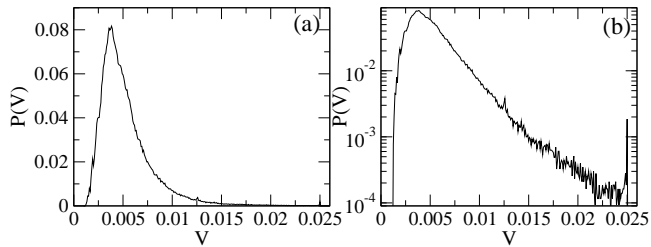}
\caption{
$P(V)$ for $\phi = 0.71$ and $F_D=0.025$, showing a peak at low $V$ near
$V = 0.037$ with a long tail extending up to $V = 0.025$. 
(b) The same $P(V)$ on a log-linear
plot indicating that the tail has an exponential form.   
 }
\end{figure}

For $\phi = 0.71$, the system is dense enough that     
the probe particle spends very little time moving freely  out
of contact with other particles.  
In Fig.~4(a) we illustrate that $V(t)$
for $\phi = 0.71$ has an intermittent character
with large velocity spikes.  
A blow up in Fig.~4(b) of one of the velocity spikes from Fig.~4(a) 
shows that there is some asymmetry in the pulses. 
A slow build up in the velocity is followed by a sharper velocity drop.     
In Fig.~5(a), $P(V)$
for $\phi = 0.71$ has a maximum near $V=0.00375$
with a long tail ending at $V = 0.025$. There
are only very small peaks near $V = 0.0125$ and $V=0.025$ 
which are remnants of the type of dynamics seen in Fig.~3(a,b). 
In Fig.~5(b) we plot $P(V)$ on a log-linear scale to indicate that
the tail has an exponential form. There are some deviations from a
single exponential fit for $ V> 0.015$ due to the existence of an upper
bound on the velocities which generates a peak at $V=0.025$.
The intermittency and asymmetry of the 
velocity jumps are very similar to the observations in 
two-dimensional drag experiments near the jamming transition \cite{Geng}. 
In those experiments, 
the stress rather than velocity fluctuations were measured, 
so the asymmetry is
reversed. The experiments of Ref.~\cite{Geng}
also showed that the fluctuations had a
maximum with an exponential
tail feature similar to that seen in Fig.~5(b) 

\begin{figure}
\includegraphics[width=\columnwidth]{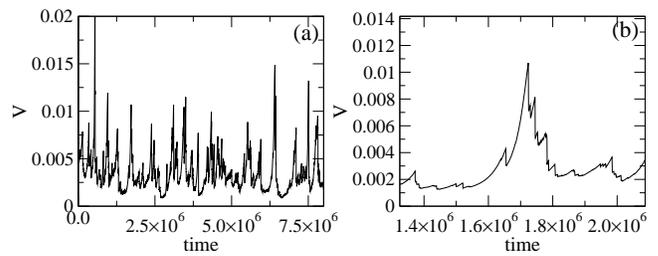}
\caption{
$V(t)$ for $\phi = 0.807$ and $F_D=0.025$, showing 
increasing intermittency. 
(b) A blowup of one of the velocity pulses showing
the same asymmetry found for $\phi = 0.71$.   
}
\end{figure}

\begin{figure}
\includegraphics[width=\columnwidth]{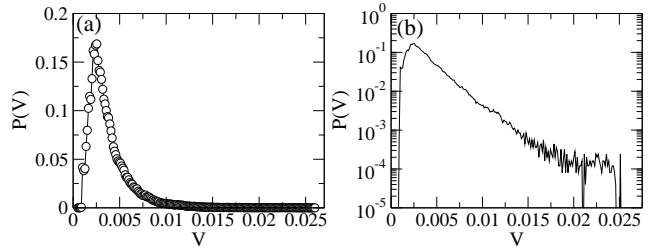}
\caption{
(a) $P(V)$ for $\phi = 0.807$ at $F_D=0.025$ 
showing a peak probability at $V = 0.0025$ with 
a long tail. (b) The same $P(V)$ on a log-linear scale,
showing that the tail fits to an exponential form.
 }
\end{figure}

In Fig.~6(a) we show $V(t)$ for $\phi = 0.807$, 
where the intermittency is more pronounced. 
The maximum value of the velocity spikes is always below $V=0.025$, 
indicating that the
probe particle is always in contact with the surrounding particles and
never moves freely.
The increasing intermittency is similar in appearance to 
the stress fluctuations seen in shear
experiments \cite{Beh}. 
Figure~6(b) shows a blowup of 
a velocity spike from Fig.~6(a) 
indicating that the same asymmetry 
observed for $\phi = 0.71$ persists. In Fig.~7(a) 
we plot $P(V)$
for $\phi = 0.807$, where the peak probability  
is shifted to a lower velocity $V=0.0025$ and the velocity tail  
is more pronounced. 
Fig.~7(b) shows $P(V)$ at $\phi=0.807$ on a 
log-linear plot illustrating that the tail has an exponential form.
In the dragging experiments of Geng and Behringer
in Ref.~\cite{Geng}, the maximum density that was studied was
around $\phi=0.77$, close to the $\phi=0.807$ density presented in 
Fig.~7. In the experimental work, the avalanche events were found 
to have an exponential distribution very similar
to the fluctuations we observe in Fig.~7(a).     
The fact that the slow drag experimental system in Ref.~\cite{Geng}
could not go to higher packing fractions 
is likely due to frictional forces, which are absent in our system.
In general we observe the same type of exponential decay in 
the tail of the velocities for 
densities up to within two percent of $\phi_{J}$.

\begin{figure}
\includegraphics[width=\columnwidth]{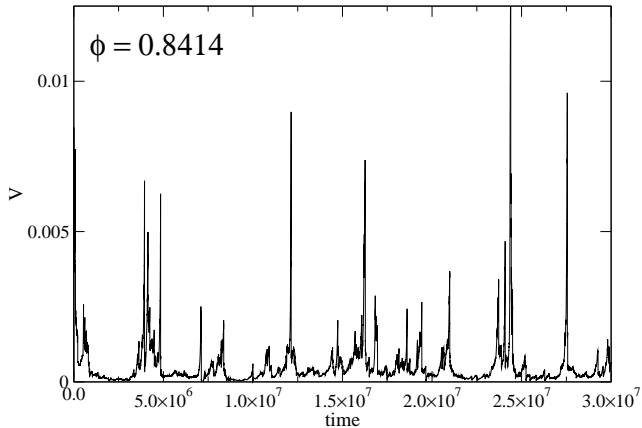}
\caption{
$V(t)$ close to jamming at $\phi = 0.8414$ and $F_D=0.025$. 
Here $V$ drops almost to zero between the velocity spikes.  
}
\end{figure}

\begin{figure}
\includegraphics[width=\columnwidth]{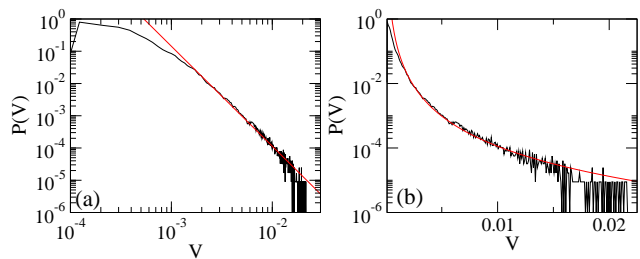}
\caption{
$P(V)$ at $\phi = 0.8414$ and $F_D=0.025$. 
(a) Log-log scale with a power law fit (light line) with $\alpha = -3.0$.
(b) The same as in (a) on a log-normal scale, 
showing that a power law fit (light line) is better than an exponential fit.  
 }
\end{figure}
    
In Fig.~8 we plot $V(t)$ for $\phi = 0.8414$, which is within 
a fraction $0.998$ of $\phi_{J}$. 
Here the fluctuations show a pronounced intermittency 
where the velocity drops almost to zero between velocity bursts. 
In Fig.~9(a) we show $P(V)$ for $\phi = 0.8414$ 
in log-log format and in 
Fig.~9(b) we show the same data on a log-normal scale. 
The light line is a a power law fit with an exponent of $\alpha = -3.0$.
The power law fit is 
valid for slightly more than a decade, so at this density we cannot 
rule out a stretched exponential form;
however, a purely exponential fit of the type performed at 
the lower densities of $\phi =0.807$ and $\phi = 0.71$ 
no longer fits the data.

\begin{figure}
\includegraphics[width=\columnwidth]{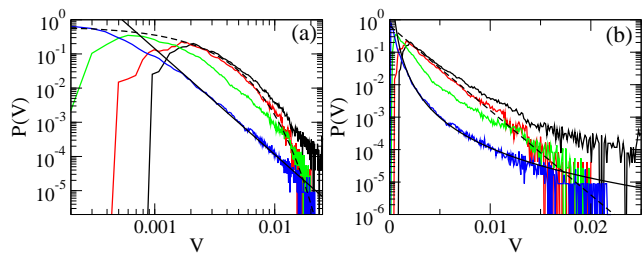}
\caption{
$P(V)$ for $\phi = 0.823$, 0.833, 0.8395, and $0.8414$ [from lower left to upper
left in panel (a)]
at $F_{D} = 0.025$. The solid line is a power law fit with $\alpha = -3.0$ and
the dotted line is an exponential fit. 
(a) Log-log scale. (b) Log-normal scale.   
 }
\end{figure}

In Fig.~10(a) we plot $P(V)$ for 
$\phi = 0.823$, 0.833, 0.8395, and  $0.8414$ on 
a log-log scale while in Fig.~10(b) we plot the same curves on a 
log-linear scale to show that
the velocity distribution tails change from an exponential to a power law
form.  The densities $\phi = 0.823$ and 0.833 are fitted well by an 
exponential, but this fit
deviates for $\phi = 0.8395$ and 
$0.8414$, crossing over to a power law form.
If point J is a critical point as suggested in recent studies, 
than the various quantities should diverge as a power  
law as $\phi_{J}$ is approached and the 
dynamical response should also reflect the
critical properties. In general, critical phenomena or fluctuations 
should only be relevant within 1 percent of the critical point or closer,
which is consistent with our observations.       
We note that the power law exponent of $\alpha=-3.0$ 
used to fit our data close to jamming is comparable to results obtained
for recent experimental studies of granular slip systems 
where the energy is released in avalanche events distributed with power law
exponents of $-2.8$ \cite{Daniels}.  

Several factors limit how closely we can approach 
$\phi_{J}$. One is the fact 
that we have a finite but small drive on the particle of 
$F_D=0.025$.  Ideally, a vanishingly small drive
would be best, but with smaller drives prohibitively 
long simulation times are necessary
to obtain adequate statistics. 
Another issue is that for densities close to $\phi_{J}$,
it possible that the probe particle can move for 
some distance and then jam. This effect was
noted in the work of O'Hern {\it et al.} 
who found that the distribution of possible jammed configurations
peaks around $\phi=0.844$ but that there was still some 
finite width to this peak as long as the system 
was finite \cite{Hern}.   

Velocity fluctuations with power law distributions have also been observed in
chute flow systems, where three phases are identified as a function of
density: a low density free-falling region, a small liquid-like region, and
a high density solid or glassy region \cite{Co}.  In the liquid-like region,
the velocities can be fit to a power law with exponents of
$-2.4$ to $-3.7$.  Even though the chute flow system is three-dimensional,
one possible interpretation of our two-dimensional results is that the
low density regime is similar to the free-falling regime, while the glassy
or solid and the liquidlike regimes could correspond to the region
just before jamming where we find power law distributions.

It is interesting to ask
why the non-exponential features such as those in Fig.~9 were not observed 
in the slow drag experiments of Ref.~\cite{Geng}. 
In the experiments the jamming density fell at a lower $\phi$ than 
our $\phi_J$, which is likely due to frictional effects. 
Although the experimental system is jammed, it does not appear to have the
same critical properties that are associated with the frictionless grains 
at $\phi_J = 0.843$. This implies that 
the concept of critical jamming may be limited to certain types of models of 
jamming and that criticality at point J 
may occur only in the frictionless limit.

\begin{figure}
\includegraphics[width=\columnwidth]{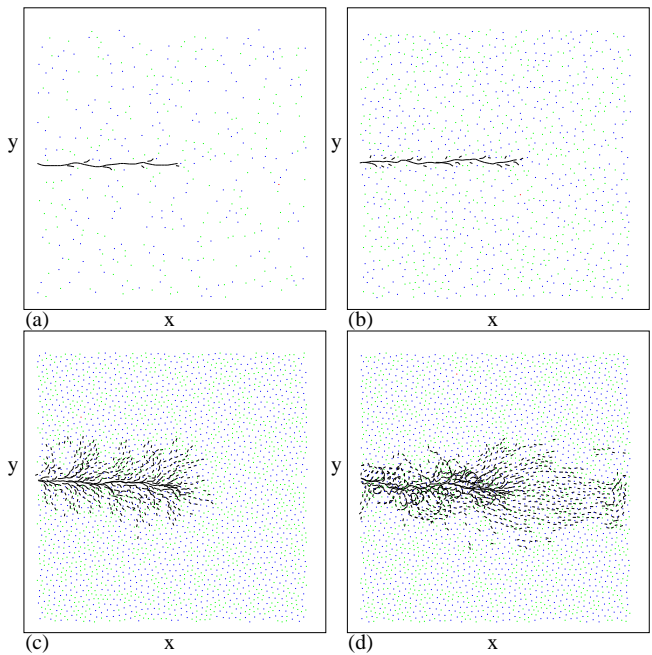}
\caption{
Black lines: trajectories followed by the particles as the probe
particle moves halfway through the system at $F_{D} = 0.025$.
(a) $\phi=0.129$. (b) $\phi=0.323$ (c) $\phi=0.807$. (d) $\phi=0.8414$.
 }
\end{figure}

In Fig.~11 we plot the trajectories of the particles as the probe moves 
halfway through the
sample. In Fig.~11(a,b) at $\phi = 0.129$ and $0.323$, 
the main trajectory of the probe particle
appears as a continuous line.  Only small perturbations of the surrounding
particles appear near the probe particle, 
showing that the perturbations are all highly localized.
For  $\phi = 0.807$ in Fig.~11(c), the perturbation extends much 
further from the probe particle in the transverse direction, 
with motion occurring for particles that are up to nine lattice constants 
away from the driven particle.
Here the trajectories of the surrounding particles are very short and do not 
intertwine.
For $\phi = 0.8414$ in Fig.~11(d),  the extent of the 
maximum transverse perturbations is only slightly larger than 
at $\phi = 0.807$; however, the perturbations ahead of the probe particle 
are much larger.
In addition, the motion of the surrounding particles takes two different
forms.  The first type of motion consists of 
those particles which move in straight
trajectories, as seen at lower $\phi$.
The second type of motion involves particles which move in larger circular 
trajectories that intertwine with the trajectories of other particles.
These trajectories are
associated with plastic deformations in which some of the surrounding
particles move around one another. 
Similar plastic rearrangements of the surrounding particles caused by the
motion of a probe particle were observed in 
a charged colloidal media \cite{Hastings}. 
In Ref.~\cite{Hastings} there was a threshold
for the motion of the driven particle, unlike the motion in Fig.~11(d) which
occurs with no threshold.
The plastic deformation trajectories that we observe
are similar to topological changes or T1 events. 
We also note that Fig.~2(b) indicated that $\langle V\rangle$ does not
develop a nonlinear dependence on $\phi$ until
$\phi > 0.825$. We only 
observe the intersecting plastic trajectories for $\phi > 0.825$. 
It is possible that the nonlinearity in $\langle V\rangle$ may be
associated with the onset of the plastic motions.

\begin{figure}
\includegraphics[width=\columnwidth]{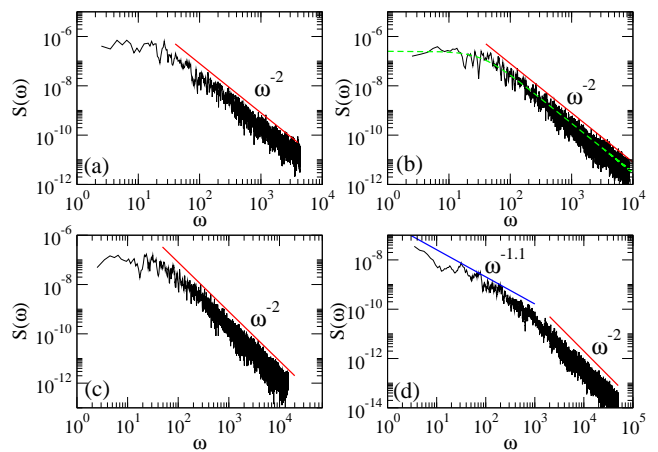}
\caption{
The power spectra $S(\omega)$ of $V(t)$ 
at constant $F_{D} = 0.025$ for (a) $\phi = 0.484$, (b) $\phi = 0.71$, 
(c) $ \phi = 0.823$
and (d) $\phi = 0.841$. For (a-c) a the high frequencies have a 
$\omega^{-2}$ form
indicated by a solid line. 
A fit to a Lorenztian is indicated by a dashed line in (b) 
and a similar fit can be performed for (a,c). 
For $\phi = 0.841$  in (d), a Lorenztian 
can no longer be fit and a low frequency
$\omega^{-1}$ feature appears.  
}
\end{figure}

\subsection{Power Spectra}

Another quantity that was measured in the slow drag experiments of 
Ref.~\cite{Geng} is the power spectrum
$$
S(\omega) = \left|\int V(t) e^{-i\omega t}dt\right|^2 . 
$$
In Ref.~\cite{Geng}, the force fluctuations showed 
an $\omega^{-2}$ behavior at high frequencies with a knee or
roll off to a flat spectrum at low frequencies.
The roll off frequency shifted to lower frequencies
with lower rotation rates. 
The same type of $\omega^{-2}$ behavior was also observed for increasing 
densities and no clear signature
appeared in the power spectrum of the fluctuations just before jamming. 
In our system, the frequency range is cut off by the inverse of the time 
required for the particle to transverse 
half of the system, so for higher densities where the probe particle
moves more slowly, we can access lower
frequencies.

For $\phi < 0.84$, we find that the power spectra of the velocity 
fluctuations have the same features 
as the experiment in Ref.~\cite{Geng}.
In Fig.~12(a,b,c) we plot the power spectra $S(\omega)$ for 
$\phi = 0.484$, 0.71, and 0.823 showing an $\omega^{-2}$
feature at high frequency and a knee feature at lower frequencies 
that is very similar to 
those seen in experiments. The power spectra can be fit to a Lorentzian form
as indicated in Fig.~12(b), 
$$
S(\omega) \propto \frac{S_{0}\omega_{0}^2}{\omega^2_{0} + \omega^2} .
$$
Here $\omega_{0}$ is the crossover frequency at the knee and $S_{0}$ is the 
spectral power along the flat portion of the spectrum below the knee.  
The presence of a Lorentzian can be interpreted as 
the emergence of a characteristic time scale.
At the lower $\phi$ the time scale could be the average time between
collisions or an average time between the large velocity bursts.  
In general, for increasing $\phi$ we find
that the knee feature shifts to lower frequencies. 

In Fig.~12(d) we plot $S(\omega)$ for $\phi = 0.841$ 
within the regime where the velocity histograms showed a power law form. 
Here the spectrum no longer has a Lorentzian form but 
shows a $1/\omega$ signal at low frequencies followed by the 
persistent $1/\omega^2$ feature at
higher frequencies.   
The appearance of $1/\omega$ noise is indicative of the presence of 
scale-free fluctuations, which are consistent with 
a critical phenomenon. 
We note that if we were able to perform very long time traces and
access lower frequencies, 
the low frequency $1/\omega$ noise power would be higher than the noise power
in the flat portion of the Lorentzian found at lower densities,
implying that the low frequency noise power $S_{0}$ 
may diverge as the jamming transition is approached. 
A divergence in the noise power at generic second order phase 
transitions has recently been proposed \cite{Yu}
which is consistent with our results. We also note that a peak 
in the noise power and the onset of $1/f$ noise
were observed two-dimensional systems at disordering transitions
\cite{Olson}.

The slow drag experiments of Ref.~\cite{Geng} 
may not have been able to access the true critical
properties of point J since the fluctuations were 
examined at a lower $\phi$ than the expected $\phi_J$.
The drag experiments of Albert {\it et al.} \cite{Albert} 
also produced Lorentzian type power spectra.
We note that this was in a three-dimensional system 
where frictional effects may also have been relevant.  

It would be interesting to see whether a crossover from Lorentzian to 
$1/f$ noise occurs experimentally
as jamming is approached. In Ref.~\cite{Fanklin}, a probe particle was pushed
from the bottom of a packing of rodlike particles and
a transition from $1/f^2$ noise 
to $1/f$ noise was correlated with a transition from 
stick-slip type motion to the motion of a solid plug.
Although this result is consistent with our results, 
the experimental system
is somewhat different in that 
there are no particle rearrangements in the solid plug state and the noise is 
due to friction effects with the wall.      

\begin{figure}
\includegraphics[width=\columnwidth]{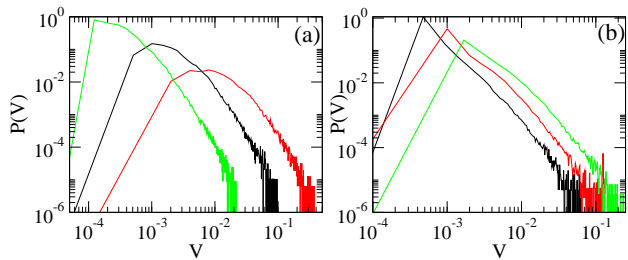}
\caption{
(a) $P(V)$ at $\phi = 0.8414$ for $F_{D} = 0.025$, 0.125, and $0.5$, 
from top left to bottom left.
(b) $P(V)$ at $\phi = 0.8426$ for $F_{D} = 0.06$, $0.125$, and $0.25$, 
from top left to bottom left.  
 }
\end{figure}

\subsection{Driving Rate Dependence}

We next examine how the velocity fluctuations vary
with the magnitude of the driving force $F_D$ on the probe particle. 
In Fig.~13 we plot $P(V)$ 
at $\phi = 0.8414$ for $F_{D} = 0.025$, $0.125$, and $0.5$.
For $F_{D} = 0.125$ there is a peak in $P(V)$ 
at $V = 0.015$  and for 
$F_{D} = 0.5$ the peak shifts to $V = 0.1$. 
In both cases the tail also shifts to higher 
$V$. For $F_{D} = 0.125$ 
the velocity tail can still be fit to a power law; however,
the range of the fit is reduced from the $F_D = 0.025$ case and for $F_D = 0.5$ 
the range of the fit is even smaller. 
This suggests that the critical fluctuations are most prominent in the
limit of small driving forces.

In Fig.~13(b) we plot $P(V)$ on a log-log scale 
at $F_D=0.06$, 0.125, and 0.25 for a system with 
$\phi = 0.8426$ which is within $0.001$ of the
jamming density. 
At this density, for $F_{D} < 0.06$
it is difficult to obtain enough statistics for a 
histogram due to the very long simulation
times; also, when the system is this close to the jamming density,
the probe particle can in some cases jam before it moves halfway 
through the system. At the jamming density the probe particle
almost aways jams instantly. 
For the lowest drive of $F_{D} = 0.06$, Fig.~13(b) shows that $P(V)$
has a power law form. 
The data can be fit with an exponent of $\alpha = -2.7$ over a range
of two decades,
while if we use only data with $ V > 0.03$,
a fit with $\alpha = -3.1$ is better.
Although we cannot more firmly establish the exact exponent, 
Fig.~13(b) indicates that the
distributions cannot be fit well to a purely exponential form.  

\begin{figure}
\includegraphics[width=\columnwidth]{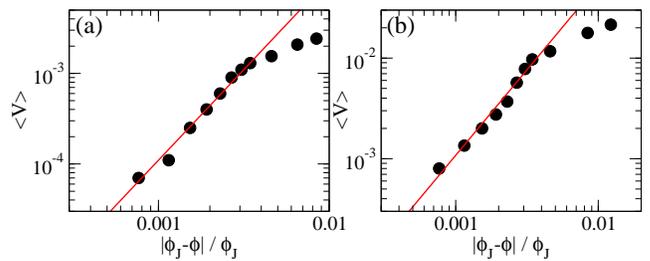}
\caption{
(a) Circles: $\langle V\rangle$ vs $|\phi_J - \phi|/\phi_{J}$ for 
$F_{D} = 0.025$. 
Straight line: power law fit with $\alpha = 2.0$.  
(b) Circles: $\langle V\rangle$ vs 
$|\phi_J - \phi|/\phi_{J}$ for $F_{D} = 0.125$.  Straight line: 
power law fit with $\alpha = 1.75$.  
 }
\end{figure}

\subsection{Estimation of the Correlation Length Scale Exponent $\nu$} 

In Ref.~\cite{Drocco}, a measurement of the number of disks
that were pushed along with the probe particle was used to
obtain the exponent of a diverging correlation length,
\begin{equation}
\xi \propto (\phi_{J} - \phi)^{-\nu} .
\end{equation} 
The resulting value of $\nu$ was between $0.6$ and $0.7$. 
Other estimates of the exponent for the same two-dimensional disk system 
have
given $\nu = 0.71 \pm 0.12$ \cite{Hern}, 
$\nu=0.6$ \cite{Olsson}, and $\nu=0.57$ \cite{Head}.
More recent quasistatic shearing simulations 
\cite{TeitelNew} have found higher estimates 
of $\nu = 0.8$ and $\nu=1.0$; it is also argued that the exponent 
is sensitive to the exact value of $\phi_{J}$, 
which may explain why the exponents in Refs.~\cite{Drocco,Olsson}
are somewhat smaller. 
References \cite{Drocco,Olsson} both used about 1200 particles, with
Ref.~\cite{Drocco} reporting 
$\phi_J=0.8395$ for a driven probe
particle and Ref.~\cite{Olsson} giving 
$\phi_J=0.8415$ for a shearing simulation.
In Ref.~\cite{Barrat}, where a higher value of $\phi_J=0.8433$ was obtained,
about 2500 particles were used.
In the present work we have a comparable number of particles $N\approx 2600$ 
and we find
$\phi_{J} = 0.843$ in agreement with Ref.~\cite{Barrat}. 
More recent shearing simulations with larger systems than in 
\cite{Olsson}
have found a higher value of $\phi_{J} = 0.843$ and $\nu = 0.8$ 
\cite{TeitelNew} so we
believe a value of $\phi_{J} = 0.843$ is reasonably accurate.   

In Fig.~2(b), the velocity $\langle V\rangle$ drops 
to zero in a nonlinear fashion as $\phi$ increases toward $\phi_J$
and the concavity in the velocity also changes very near jamming.   
We can estimate $\nu$ based on the scaling behavior of $\langle V\rangle$ as
the jamming transition is approached.
At $\phi=0$, the probe particle moves with a free velocity of $V_0$.  At
nonzero $\phi$, the velocity of the probe particle decreases according to
$V \propto V_0/N_m$, where $N_m$ is the number of other disks that are being
pushed along by the probe particle.
If the pushed particles are within a disklike jammed area 
of radius $\xi$ that surrounds the
probe particle, then $N_m=\phi_J\xi^{2}$.
If $\xi \propto (\phi_{J} - \phi)^{-\nu}$, then 
the velocity should scale as  
$V \propto V_{0}(\phi - \phi_{J})^{2\nu}$. 
From the velocity versus $\phi$ curves we can fit 
$V \propto (\phi - \phi_{J})^\alpha$, 
giving $\nu = \alpha/2$. In Fig.~14(a) we plot 
$V$ vs $|\phi_J -\phi|/\phi_J$ 
for $F_{D} = 0.025$ along with a fit of $\alpha=2.0$, 
which gives $\nu = 1.0$. 
Figure 14(b) shows the same plot for $F_D=0.125$ where we obtain
$\alpha=1.75$ and $\nu=0.875$.
These values are within the range of the estimates from   
the quasistatic shearing simulations of Ref.~\cite{Barrat}. 
We note that  other theoretical studies have found $\nu = 0.5$  
\cite{Silke,Wyart1,El}. 
Although our scaling range is too small to give a definitive value for $\nu$, 
our results along with the other simulations 
\cite{Drocco,Olsson,Barrat,Head} are consistent with
$\nu > 0.5$. 
It is possible that there is more than one diverging length scale near
jamming.
From the images of the particle displacements near jamming 
in Fig.~11, there appear to be
two different modes of displacements of 
the surrounding particles: the small displacements
that extend in front of the driven particle and the plastic deformations 
that occur further from the probe particle.
It may be that there is a different length scale associated with each of
these two types of motion.

\begin{figure}
\includegraphics[width=\columnwidth]{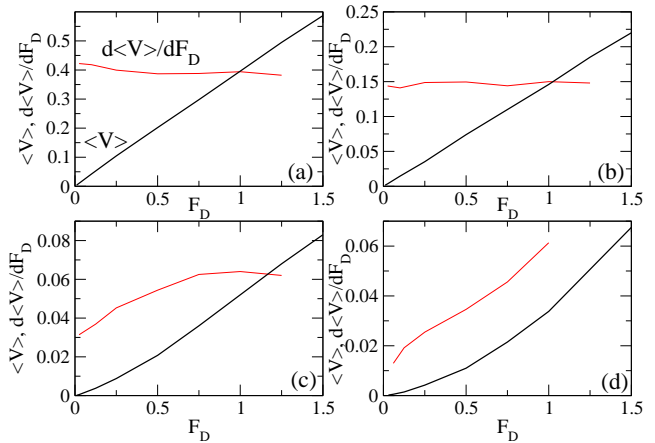}
\caption{
$\langle V\rangle$ vs $F_{D}$ (dark curves) and 
$d\langle V\rangle/dF_{D}$ vs $F_{D}$ (light curves).
The velocity-force curves are linear for (a) $\phi = 0.483$ and 
(b) $\phi = 0.807$
and are nonlinear for (c) $\phi = 0.8414$ and  (d) $\phi = 0.8426$.    
 }
\end{figure}

\subsection{Velocity Force Curves}

We next examine the average velocity $\langle V\rangle$ versus $F_{D}$ 
for varied $\phi$.  
At $\phi=0$ in the absence of other disks, the overdamped dynamics we employ
cause the velocity of the probe 
particle to vary linearly with the applied force, $V \propto F_{D}$. 
For $\phi$ 
well below  $\phi_{J}$, 
collisions with other particles reduce the probe particle velocity and
increase the effective viscosity; 
however, the velocity-force curves remain linear as shown 
in Fig.~15(a,b) for $\phi = 0.483$ and $\phi=0.807$. 
The plots of $d\langle V\rangle/dF_{D}$ in Fig.~15(a,b) have little curvature,
which is consistent with a linear velocity-force relationship.
Very close to jamming the velocity-force curves become nonlinear  
as seen in Fig.~15(c,d) for $\phi = 0.8414$ and $\phi=0.8426$, where the 
$d\langle V\rangle/dF_{D}$ curves show growing deviation from linearity.  

\begin{figure}
\includegraphics[width=\columnwidth]{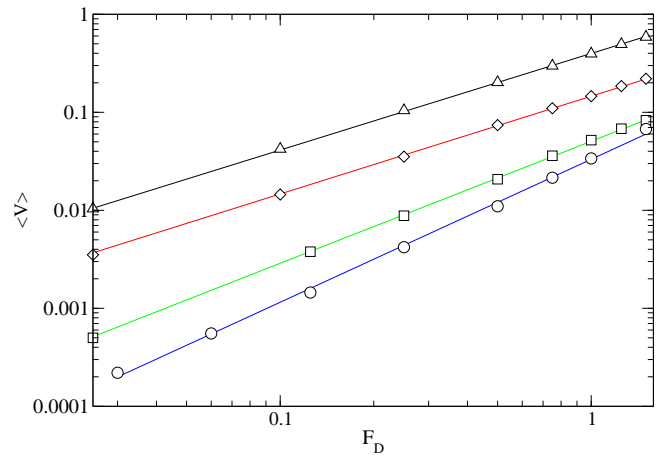}
\caption{
$\langle V\rangle$ vs $F_{D}$ on a log-log scale for 
$\phi = 0.483$ (triangles), $\phi=0.807$ (diamonds), $\phi=0.8414$ (squares), 
and $\phi=0.8426$ (circles). 
The straight lines are power law fits with 
$\beta = 1.0$ for $\phi = 0.483$ and $0.807$, 
$\beta=1.24$ for $\phi=0.8414$, and $\beta=1.47$ for $\phi=0.8426$.}
\end{figure}

In Fig.~16 we plot $\langle V\rangle$ vs $F_{D}$ 
on a log-log scale where the solid lines are
power law fits with the form $V = F_D^\beta$. 
For $\phi = 0.483$ and $\phi=0.807$ the fits are linear
with $\beta=1$, while
for $\phi = 0.8414$ and $\phi=0.8426$, 
the fits give $\beta = 1.24$ and $\beta=1.47$, respectively. 
In general, exponents are not expected to change 
continuously near a critical point so the
exponent at $\phi = 0.8414$ probably originates from a smeared 
combination of linear and nonlinear regions. The
exponent of $\beta=1.47$ at $\phi=0.8426$ 
should be more accurate and is also in agreement with a number of 
previous studies of single driven probe particles. 
Simulations of a driven particle 
moving through a disordered assembly of 
charged colloids gave an exponent of $\beta=1.5$ \cite{Hastings,LM}. 
In both the colloidal system and the disk system, there are plastic
rearrangements of the particles around the probe particle; however, there are
several differences between the two systems, including the fact that the 
colloidal particle is effectively 
jammed at low drives and has a nonzero critical
threshold force for motion.
Additionally, 
in the charged colloid system there is no well defined
jamming density $\phi_{J}$ due to the longer range soft Yukawa 
interaction potential.    
Experiments where a single probe particle is driven through a 
three-dimensional colloidal system as the density is increased towards the
glass transition gave an exponent of $\beta=1.5$ at densities below the
glass transition, which can be interpreted as below the jamming transition
\cite{Weeks}.
The closest experimental realization of our system is the work in Ref.~\cite{CD}
where a probe particle is driven through a two-dimensional assembly of grains. 
One difference between the experiments in 
Ref.~\cite{CD} and our simulations is that the experiment contains
an additional vibration to induce a   
fluidized state. 
In the experiment, linear velocity-force curves appear at low densities, while
closer to jamming there is a nonlinear region at low drives with a 
crossover to a linear region at high
drives. 
This crossover shifts to higher values of $F_{D}$ as 
the jamming transition is approached.     
In the nonlinear regime, 
Ref.~\cite{CD} used a fit to $V \propto \exp(F_D)$ 
rather than a power law; however, the range of the data is sufficiently
small that it does not rule out a power law fit.
Additionally, it is possible that the applied vibrations
induce activation dynamics that are not present in our work. 
Despite these differences,
the onset of nonlinear velocity-force curves and increased intermittency 
close to jamming found in Ref.~\cite{CD} are consistent with our results.  

We note that a power law fit to the velocity force curves has been associated
with criticality at the onset of motion in other two-dimensional systems.   
In particular, extensive studies have been performed on
plastic depinning of particle assemblies driven over
rough landscapes in systems ranging from
colloidal particles moving over disordered substrates \cite{Reichhardt2,Ling},
vortices in type-II superconductors \cite{Dominguez},
and generalized models of particle 
motion over quenched disorder \cite{Fisher}.  
In these systems, the velocity response  can be fit to the form
\begin{equation}
V = (F_{D} - F_{C})^{\beta}.
\end{equation}  
Here $F_{C}$ is the threshold force that must be applied for motion to occur. 
For $\phi  < \phi_{J}$, we find $F_{C} = 0$; 
however, velocity-force curves
with nonlinear scaling can still occur even in the absence of a threshold.    
In the collective depinning systems, exponents of $\beta=1.5$ to 
$\beta=2.25$ are generally observed. 
Our results showing a crossover from linear to nonlinear
velocity-force curves near jamming are 
a further indication that point J exhibits critical properties
and that this critically also appears in transport. 

\begin{figure}
\includegraphics[width=\columnwidth]{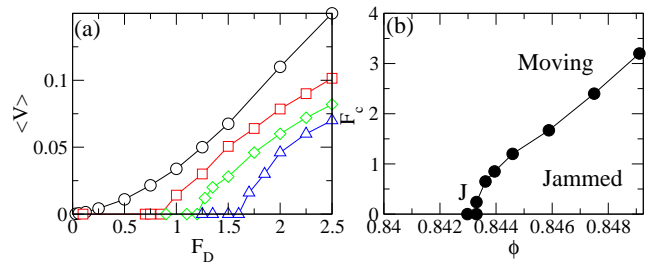}
\caption{
(a) $\langle V\rangle$ vs $F_{D}$ 
below jamming at $\phi = 0.8426$ (circles) and above jamming at
$\phi = 0.8439$ (squares), $\phi = 0.8446$ (diamonds), 
and $\phi = 0.846$ (triangles).
Above jamming, a threshold force $F_{c}$ 
is needed to depin the probe particle.    
(b) The threshold force $F_{c}$ vs 
$\phi$ with point $J$ denoting where the system
becomes jammed.}   
\end{figure}

\section{Threshold and Yielding Above Jamming} 

We next show that for $\phi > \phi_{J}$, there is a finite threshold 
force required to move the probe particle through the background of disks.
In the jammed phase, it is known that the onset of motion 
or yielding consists
of a coexistence between an unjammed fluid phase and a 
jammed solid phase, and that
in the presence of a shear, the motion may be confined within a certain
region due to shear banding \cite{Dennin}. 
In Fig.~17(a) we 
plot $\langle V\rangle$ vs $F_{D}$ for $\phi = 0.8426$, 0.8439,
0.8446, and $0.846$.
Here the probe is 
immobile or pinned at low drives for $\phi > \phi_{J}$ 
as indicated by the fact that  $\langle V\rangle = 0.0$
until the external force $F_D$ reaches a critical force $F_c$
for motion.
At densities just below jamming, Fig.~16(b) showed that the velocity-force
curves have positive curvature with $\beta>1$.
Above jamming, the
velocity-force curves have a different form and 
are more consistent with either a linear dependence $\beta=1$ or with
$\beta < 1.0$. Although we do not have the accuracy in this regime to determine
the exact form of the velocity-force curves above jamming, 
our results suggest that the
onset of motion is different for $\phi > \phi_{J}$ than for
$\phi<\phi_J$.       

The threshold
force $F_c$ increases with increasing $\phi$. In Fig.~17(b) we plot $F_{c}$   
vs $\phi$ showing the onset of the threshold at point $J$ and 
the subsequent rapid increase of $F_c$
with increasing $\phi$. 
Well above $\phi_J$, $F_c$ increases roughly linearly with increasing $\phi$.
The appearance of a threshold and the general shape
of the $F_{c}$ curve is in agreement with recent three-dimensional colloid
simulations 
in which a threshold for motion appears above a certain packing density 
\cite{Ga}.      
We note that in our system, the disks interact with a harmonic potential 
and are soft enough that the probe particle can squeeze between neighboring
disks above jamming.  A similar yielding
appears in shearing simulations performed above jamming \cite{Olsson,Barrat}.
For real granular systems with frictional effects and 
much stiffer interactions between 
particles, it would be difficult to move the probe particle through 
the system above jamming. For colloidal
particles with much softer interactions, 
a threshold force can be extracted as has been demonstrated in 
experiment \cite{Weeks}. 

\begin{figure}
\includegraphics[width=\columnwidth]{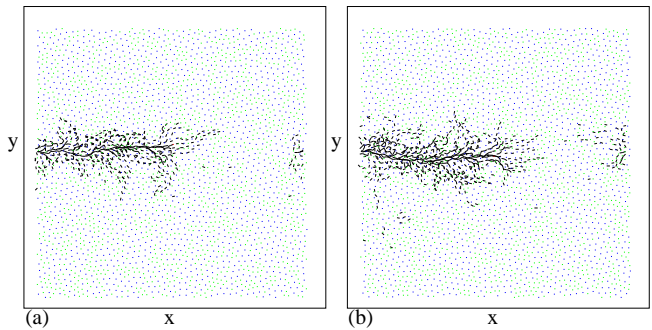}
\caption{
Particle trajectories during the time period required for the probe particle 
to move halfway through the system at a drive just above the depinning
threshold $F_c$.
(a) $\phi = 0.8439$. (b) $\phi = 0.846$.}   
\end{figure}

In Fig.~18(a,b) we plot the particle trajectories at a drive
just above the depinning threshold for    
$\phi =0.8439$ and $\phi = 0.846$, where plastic deformations occur within
a localized region.
The trajectories are qualitatively similar to the case 
$\phi = 0.8414$ shown in Fig.~11(d) where the
motion 
consisted of a combination of particles close to the driven particle being
pushed linearly over a small distance while particles further from the
driven particle undergo intertwining plastic motion.
One noticeable difference is that the spatial extent of the perturbations above
jamming is less than that seen 
below jamming at $\phi = 0.8414$. Additionally, the
spatial extent of the perturbations above jamming 
does not appear to rapidly grow in size 
as $\phi$ increases above $\phi_J$.
The fact that the motion is more localized 
above jamming suggests that 
in this regime the system is only plastically distorted near the probe 
while the remaining portion of the system acts rigidly elastic. 
We call this a local shear banding effect, and it suggests that 
the onset of motion or yielding above jamming 
is consistent with the behavior of a disordered 
elastic solid. A similar conclusion that the state above 
jamming is a disordered
elastic solid was drawn from two-dimensional shearing simulations \cite{Barrat}.      

\begin{figure}
\includegraphics[width=\columnwidth]{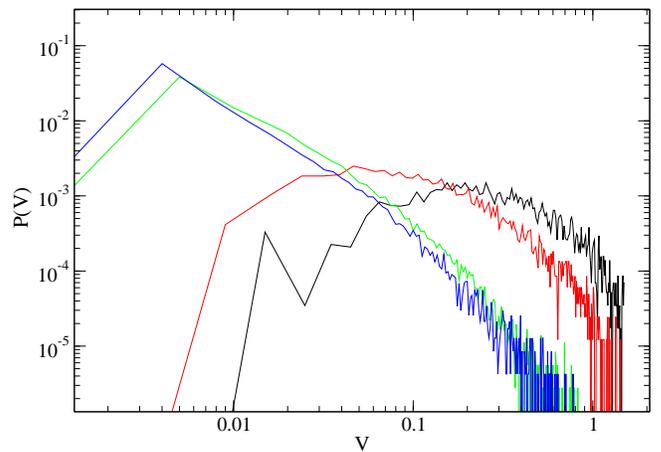}
\caption{Velocity histograms $P(V)$ above jamming at $\phi= 0.846$ for 
$F_{D} = 1.725$, 1.85, 4.0, and $6.0$, from upper left to lower left.} 
\end{figure}

In Fig.~19 we plot $P(V)$ for $\phi = 0.846$ at 
$F_{D} = 1.725$, 1.85, 4.0, and $6.0$.
At $F_{D} = 1.725$ and $F_D=1.85$, 
$P(V)$ has a broad form with non-exponential features.
Attempts to perform a single power law 
fit for the higher values of $V$ in this regime give only a limited range 
with an exponent of $-3.0$.      
For the higher drives $F_D=4.0$ and $F_D=6.0$,
there is still a tail at high drives; however, 
a power law fit cannot be performed.

\begin{figure}
\includegraphics[width=\columnwidth]{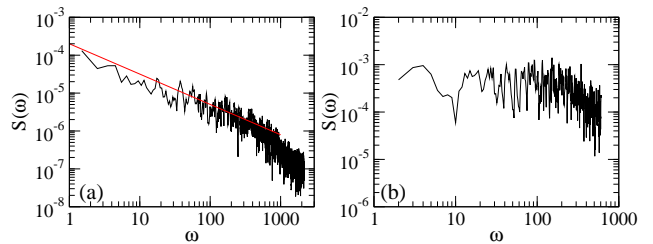}
\caption{The power spectra $S(\omega)$ of the velocity time traces $V(t)$
at $\phi = 0.846$. (a) $F_{D} = 1.725$, just above depinning. The solid line
is a power law fit with $\alpha = -0.8$ (b) $F_{D} = 6.0$. A flat
spectral signature appears at low frequencies.} 
\end{figure}

In Fig.~20(a) we show the power spectrum $S(\omega)$ 
of $V(t)$ at $\phi=0.846$ for $F_{D} = 1.725$, just above depinning.
The solid line is a power law fit to
$S(\omega) \propto \omega^{-\alpha}$.  We obtain $\alpha = 0.8$ over a      
wide range of low frequencies, with a  
small region of $\omega^{-2}$ dependence at high frequencies.
The smallness of the $\omega^{-2}$ regime is a reversal from what
we observe below jamming,
where the $\omega^{-2}$ frequency dependence
dominates over a large range until just below the jamming
transition. In Fig.~20(b) we 
plot $S(\omega)$ for the same density 
$\phi=0.846$ but at $F_{D} = 6.0$, well above the depinning
threshold. In this case, 
at low frequencies a white spectrum with $\alpha = 0$ appears.

\section{Summary} 
In summary, we have  numerically investigated the velocity 
fluctuations of a driven probe particle as it moves through
a bidisperse assembly of disks in two dimensions. 
The particular model we study is composed of harmonically interacting 
disks with a $1:1.4$ radius ratio. 
We find a jamming density of $\phi_{J} = 0.843$, in agreement with
recent numerical work on the same system \cite{CD,Head,Barrat}. 
For a constant driving force 
on the probe particle, we find 
that the average probe velocity goes to zero continuously
as $\phi$ approaches $\phi_{J}$ from below 
and that the velocity fluctuations develop increasingly 
intermittent features close to $\phi_J$.
The velocity distribution functions 
have long tails which fall off exponentially, similar to the
force fluctuations observed in the slow drag experiments of 
Geng and Behringer \cite{Geng}. When $\phi$ is within one percent of 
$\phi_{J}$,
the velocity distributions can no longer be fit to a purely exponential
form and have a power law shape suggesting that the
system is exhibiting fluctuations of a critical type.
The extent of the perturbations caused by the driven grain in the surrounding
packing also grows as jamming as approached and the perturbations take two
forms.
The first is small linear displacements of less than a lattice constant, while
the second is large winding displacements which are 
associated with plastic rearrangements. 
The plastic rearrangements occur only very near point J 
where the power law velocity distributions are observed.    
The power spectrum of the fluctuations 
below jamming has a Lorentzian shape with a knee at a crossover frequency
and $\omega^{-2}$ behavior at higher frequencies, similar to the
spectral signatures found in slow drag experiments \cite{Geng}.
Just below the jamming transition, the power spectrum 
changes shape and develops a $1/\omega$ low frequency behavior.
The velocity-force curves are linear below jamming, but become 
nonlinear with a power law form just at the jamming transition.
The appearance of nonexponential velocity distributions, 
the onset of $1/\omega$ noise and the development of
power law velocity-force curves
are all consistent with jamming having the signatures of
a critical phenomenon.    
From the velocity versus $\phi$ curves, 
we obtain an estimate for the correlation length exponent 
$\nu = 0.8-1.0$, in agreement 
with recent studies \cite{Barrat,TeitelNew}.   
For $\phi > \phi_{J}$ we find that a critical threshold force $F_{c}$ 
is required to set the probe particle in motion,
in agreement with recent numerical studies of driven probe 
particles in dense colloidal assemblies
at densities above  jamming \cite{LM}.  
The threshold force increases with increasing 
$\phi$ and the motion consists of 
localized plastic distortions near the moving particle 
which we call a local shear banding effect.      
The velocity fluctuations at drives just above the 
threshold force have a nonexponential form 
and the power spectrum exhibits a $1/\omega$ shape. 
For higher drives, the velocities are exponentially distributed, 
suggesting that the onset of yielding by the 
probe particle for $\phi > \phi_{J}$ has similar properties to 
yielding at point $J$.   

We thank M. Hastings, L. Silbert, and S. Teitel for useful comments.    
This work was carried out under the auspices of the NNSA of the U.S. DoE
at LANL under Contract No. DE-AC52-06NA25396.

\end{document}